# Fonctions de correction de la diffraction et solutions analytiques dans les mesures en acoustique non linéaire

## Functions of diffraction correction and analytical solutions in nonlinear acoustic measurement

L. Alliès, D. Kourtiche, M. Nadi

L.I.E.N, Université H. POINCARE, BP 239, 54506 Vandœuvre, France.

**Summary**
This paper presents an analytical formulation for correcting the diffraction associated to the second harmonic of an acoustic wave, more compact than that usually used. This new formulation, resulting from an approximation of the correction applied to fundamental, makes it possible to obtain simple solutions for the second harmonic of the average acoustic pressure, but sufficiently precise for measuring the parameter of nonlinearity B/A in a finite amplitude method. Comparison with other expressions requiring numerical integration, show the solutions are precise in the nearfield.

**Sommaire**
Dans cet article nous présentons une formulation analytique de la correction de la diffraction appliquée au second harmonique de la pression acoustique, plus compacte que celle habituellement utilisée. Cette nouvelle formulation, obtenue à l'aide d'une approximation de la correction appliquée au fondamental, permet d'obtenir des solutions simples pour le second harmonique de la pression acoustique moyenne, mais suffisamment précises pour être utilisées dans des systèmes de mesure du paramètre de non-linéarité B/A utilisant les méthodes d'amplitude finie. Des simulations effectuées dans le champ proche montrent la bonne précision de ces solutions par rapport à celles qui nécessitent une intégration numérique.

## 1. Introduction

Dans les mesures des paramètres acoustiques d'un milieu, effectuées en transmission directe ou en mode pulse-écho, il est nécessaire de prendre en compte les effets de la diffraction de la source ultrasonore pour améliorer la précision des mesures.
Les cellules de mesure couramment utilisées sont constituées soit de deux transducteurs plans de forme circulaire, l'un servant de source et l'autre de détecteur, soit d'un seul transducteur utilisé en émetteur-récepteur pour le mode pulse-écho. Dans ces situations le transducteur détecteur traduira en tension électrique la pression acoustique moyenne exercée sur sa surface de réception. Et que l'on travaille dans le cadre de l'acoustique linéaire ou non linéaire, les solutions analytiques décrivant cette pression moyenne peuvent se formuler comme la somme de deux termes, l'un correspondant à la propagation d'une onde plane, et l'autre intégrant les effets de la diffraction engendrés par la géométrie de l'ensemble source-détecteur. Les fonctions de diffractions apparaissent donc comme des corrections à apporter à la propagation d'une onde plane théorique pour l'adapter à une situation réelle.
Les principaux paramètres acoustiques caractérisant un milieu sont l'atténuation $\alpha$, la vitesse de phase *c*, et le paramètre de non-linéarité *B/A*. Les deux premiers paramètres se mesurent dans le cadre de l'acoustique linéaire, c'est à dire à partir de la solution analytique issue de l'approximation linéaire de l'équation générale de propagation. On parle dans ce cas d'onde d'amplitude infinitésimale.
Dans ce cadre, différents auteurs [1, 2, 3, 4] ont donné des expressions, exacte et asymptotiques, de la pression moyenne reçue par un transducteur de surface circulaire placé dans le même axe que celui d'une source de géométrie identique. Ces expressions permettent d'établir des fonctions de correction de la diffraction dans les mesures de vitesse de phase et d'atténuation [5, 6].
Par contre le paramètre B/A se mesure dans le cadre de l'acoustique non linéaire, et on distingue 2 familles de méthode de mesure :

- *Les méthodes thermodynamiques*, qui sont déduites de la définition même du paramètre B/A dans l'équation d'état. Si elles sont les plus précises, elles nécessitent un appareillage complexe [7, 8].
- *Les méthodes d'amplitude finie,* dans lesquelles le paramètre B/A est déduit de façon indirecte en quantifiant la distorsion d'une onde ultrasonore se propageant dans un milieu non linéaire. Dans ce cas on utilise généralement des solutions analytiques asymptotiques issues de l'approximation quasi linéaire de l'équation de propagation. Dans cette approximation on néglige les harmoniques d'ordre supérieur à 2 générés dans le milieu par une onde initialement sinusoïdale, et on considère que l'harmonique 2 n'engendre pas de décroissance appréciable du fondamental par transfert d'énergie.

Les premières mesures du paramètre B/A par les méthodes d'amplitude finie reposaient sur une expression analytique du second harmonique en considérant la propagation d'une onde plane [9, 10, 11, 12]. Différents auteurs [13, 14, 15, 16, 17] ont ensuite amélioré la précision de ces méthodes en incluant une fonction de correction de la diffraction issue de la formulation établie par Ingenito et Williams [18] pour la pression moyenne exercée par le second harmonique. Cependant, la correction de la diffraction obtenue n'est pas très pratique car elle ne peut être évaluée que par une intégration numérique.

L'objectif de cet article est de montrer que l'on peut obtenir une forme beaucoup plus simple, mais suffisamment précise, en simplifiant la correction de la diffraction pour le fondamental dont est issue celle du second harmonique. Puis nous donnerons des expressions simples de la pression moyenne exercée par le second harmonique, incluant les effets de la diffraction et de l'atténuation, et nous montrerons qu'elles sont confondues avec celle établie par Coob et vérifiée expérimentalement [13].
Mais avant d'établir ce résultat il est nécessaire de présenter les différentes corrections de la diffraction applicables au fondamental de la pression acoustique.

## 2. Correction de la diffraction et pression acoustique moyenne du fondamental

### 2.1. Solution de l'équation de propagation

Pour une onde monochromatique de nombre d'onde $k=\omega/c_o$, le fondamental est gouverné par l'équation de Helmholtz en milieu dissipatif :

$$\Delta\phi_1(1-2j\alpha_1/k)+k^2\phi_1=0 \qquad (1)$$

$\phi_1$ représente le potentiel des vitesses particulaires du fondamental lié à la pression $p_1$ par $p_1 = -j\rho_o\omega\phi_1$. $\rho_o$ est la masse volumique du milieu, $\alpha_1$ son atténuation à la fréquence $f$ du fondamental, et $c_o$ la vitesse de propagation de l'onde ultrasonore.
Pour une source plane de rayon $a$ vibrant avec une amplitude $U_0$, et avec la condition $\alpha_1 << k/2$, la solution exacte de l'équation (1) en coordonnées cylindriques *(r,z)* est l'intégrale de King [19] :

$$\phi_1(r,z) = U_0 a \int_0^\infty J_0(\psi r) J_1(\psi a) \frac{e^{jz(k'^2-\psi^2)^{1/2}}}{(k'^2-\psi^2)^{1/2}} d\psi \quad \text{avec} \quad k' = k + j\alpha_1 \qquad (2)$$

2.2. Potentiel et pression acoustique moyenne du fondamental

Conformément à la configuration géométrique (figure 1), le potentiel moyen $<\phi_1(r,z)>$ sur la surface du transducteur détecteur est :

$$\langle\phi_1(r,z)\rangle = \frac{\int_S \phi_1(r,z) dS}{\pi a^2} = \frac{2\int_0^a \phi_1(r,z) r dr}{a^2} \qquad (3)$$

z est la distance séparant le détecteur de l'émetteur, et le choix d'un rayon identique pour les deux transducteurs conduit à des simplifications théoriques pour le potentiel moyen.

A partir de l'expression (2) pour le cas non dissipatif *($\alpha_1 = 0$)*, Williams [1] a donné l'expression exacte du potentiel moyen :

$$\langle\phi_1(r,z)\rangle = \frac{jU_0}{k} e^{jkz} - \frac{j4U_0}{k\pi} \int_0^{\pi/2} e^{jk[z^2+4a^2\cos(\theta)^2]^{1/2}} \sin(\theta)^2 d\theta \qquad (4)$$

Le premier terme représente le potentiel des vitesses dans le cas d'une onde plane, donc le potentiel moyen sur la surface de réception : $\phi_{10}(z) = \frac{jU_0}{k} e^{jkz} = \langle\phi_{10}(r,z)\rangle$. L'expression (4) met ainsi en évidence la décroissance du potentiel moyen par rapport à une onde plane théorique, engendrée par les effets de diffraction du système émetteur-détecteur traduit par le second terme. Et la pression acoustique moyenne exercée sur le détecteur s'exprime sous la forme : $\langle p_1(r,z)\rangle = -j\rho_o\omega\langle\phi_1(r,z)\rangle$

2.3. Fonction $D_1(z)$ de correction de la diffraction pour le fondamental

La fonction de correction de la diffraction $D_1(z)$ permet d'adapter l'onde plane théorique à une situation réelle. Par conséquent:

$$D_1(z) = \frac{\langle\phi_1(r,z)\rangle}{\langle\phi_{1o}(r,z)\rangle} = \frac{\langle p_1(r,z)\rangle}{\langle p_{1o}(r,z)\rangle} \quad \text{avec} \quad \langle p_{10}(r,z)\rangle = P_0 e^{jk'z} \qquad (5)$$

$<P_{10}(r,z)>$ représente la pression moyenne exercée par le fondamental dans le cas d'une onde plane, et $P_0 = \rho_0 c_0 U_0$ est la pression acoustique moyenne à la source. Ainsi le module de la pression moyenne s'exprime en milieu dissipatif sous la forme :

$$|\langle p_1(r,z)\rangle| = P_0 e^{-\alpha_1 z} |D_1(z)| \qquad (6)$$

L'expression exacte de $D_1(z)$ est obtenue avec la solution de Williams (4) :

$$D_1(z) = 1 - \frac{4}{\pi} e^{-jkz} \int_0^{\pi/2} e^{jk[z^2 + 4a^2 \cos(\theta)^2]^{1/2}} \sin(\theta)^2 d\theta \quad (7)$$

2.4. Simplifications de la fonction de correction $D_1(z)$

Pour $z > a$ Bass [2] a donné une très bonne approximation de la solution (7)[1] :

$$D_1(z) \approx 1 - \left(1 - \frac{\xi(z)^2}{2(ka)^2}\right)\left(J_0(\xi(z)) + jJ_1(\xi(z))\right)e^{-j\xi(z)}$$

$$- \left\{\frac{\xi(z)}{(ka)^2} J_1(\xi(z))e^{-j\xi(z)}\right\} \quad \text{avec} \quad \xi(z) = \frac{k}{2}\left(\sqrt{z^2 + 4a^2} - z\right) \qquad (8)$$

Sous la condition $1 < z/a << ka$, ce qui implique $\xi(z) >> 1$, on peut utiliser les développements asymptotiques des fonctions de Bessel et négliger le terme entre { } de (8). Soit :

$$D_1(z) \approx 1 - \left(1 - \frac{\xi(z)^2}{2(ka)^2}\right)\left(\frac{2}{\pi\xi(z)}\right)^{\frac{1}{2}} e^{-j\frac{\pi}{4}} \qquad (9)$$

Cette expression conduira à la simplification (24) sur $D_1(z)^2$ exploitée par Coob [13] pour évaluer la pression moyenne du second harmonique.

En limitant au 1er ordre le développement de $[\ ]^{1/2}$ dans la solution (7), Rogers *et al* [4] ont obtenu une bonne approximation sous la forme :

$$D_1(z) \approx 1 - e^{-j\frac{ka^2}{z}} \left[J_0\left(\frac{ka^2}{z}\right) + jJ_1\left(\frac{ka^2}{z}\right)\right] \qquad (10)$$

Elle est valide pour toutes les valeurs de $z/a$ si $(ka)^{1/2} >> 1$, et l'erreur apportée par cette simplification par rapport à la solution exacte (7) est inférieure à 0.4 % pour ka = 100.

Notons que cette relation peut être obtenue à partir de l'expression (8) en limitant au 1er ordre le développement de la fonction $\xi(z)$ et en négligeant les termes en $1/(ka)^2$.

Pour $z/a < (ka)^{1/2}$, la condition précédente implique $ka^2/z > (ka)^{1/2} >> 1$, et on peut réduire l'expression (10) en utilisant les développements asymptotiques des fonctions de Bessel :

$$D_1(z) \approx 1 - \left(\frac{2.z}{\pi.k.a^2}\right)^{1/2} .e^{-j.\pi/4} \qquad (11)$$

Si $z/a >> 1$, le développement de la fonction $\xi(z)$ peut être limité au 1er ordre, soit $\xi(z) \approx ka^2/z$, et avec la condition $(ka)^{1/2} >> 1$ la solution (9) converge vers la solution (11) comme le montre la figure 2.c. Le domaine de validité de la solution (9) peut donc être défini comme celui de la solution (11), c'est à dire par $z/a < (ka)^{1/2}$ pour $(ka)^{1/2} >> 1$.

---

[1] *Corrigée par Williams [4].*

D’une façon générale la fonction de correction peut se mettre sous la forme :

$$D_1(z) = 1 - g(z) \quad (12)$$

où nous définissons *g(z)* comme la fonction de diffraction, liée aux paramètres de source *a* et *k*, et possédant la propriété $\lim_{ka \to \infty} [g(z)] = 0$ (cas d'une onde plane).

2.5. Comparaison des différentes expressions de $D_1(z)$

Les figures 2.a et 2.b représentent le module $|D_1|$ des différentes expressions de la fonction de correction à l'aide des variables *z/a* et $s = z\lambda/a^2 = 2\pi z/ka^2$, cette dernière permettant de distinguer le champ proche $(s \leq 1)$ du champ lointain $(s > 1)$. Les simulations sont effectuées pour $a = 1$ *cm* et $ka = 125$.
Les simplifications (8-10) sont confondues avec la solution exacte (7), et les expressions asymptotiques (9-11) constituent de bonnes approximations dans le champ proche (fig. 2.a), elles divergent de la solution exacte pour $z/a > 60$ soit $s > 3$ (fig. 2.b). L'erreur relative (fig. 2.c) met en évidence la convergence de la solution (9) vers la solution (11) et confirme le domaine de validité $z/a < (ka)^{1/2}$, ou $s < 2\pi/(ka)^{1/2}$, pour $(ka)^{1/2} >> 1$. Dans ce domaine l'erreur relative est inférieure à 0.7 % excepté au voisinage immédiat de la source pour la solution (9), la borne inférieure étant de toute façon limitée expérimentalement à l'apparition d'ondes stationnaires dans la cellule de mesure.

## 3. Correction de la diffraction et pression acoustique moyenne du 2ème harmonique

A partir de la formulation du potentiel moyen du second harmonique établie par Ingenito et Williams [18], nous montrons dans cette partie que l'on peut exprimer sous une forme compacte la fonction de correction de la diffraction pour le second harmonique $D_2(z)$ en fonction de celle pour le fondamental $D_1(z)$.

3.1. Solution de l'équation de propagation

Sur la base de l'approximation quasi linéaire appliquée aux équations de l'hydrodynamique [20], Ingenito et Williams [18] dérivent une équation gouvernant le second harmonique pour une onde monochromatique en milieu non dissipatif :

$$\Delta\phi_2 + 4k^2\phi_2 = -j\frac{k^3[1 + (B/A)/2]}{c_0}\phi_1^2 \quad (13)$$

$\phi_2$ représente le potentiel des vitesses du second harmonique lié à la pression $p_2$ par $p_2 = -2j\rho_o\omega\phi_2$ , et *B/A* est le paramètre de non-linéarité. Ingenito et al ont donné une bonne approximation de la solution de cette équation, qui peut être adaptée au cas dissipatif [13] sous la forme :

$$\phi_2(r,z) \approx \frac{-\beta k^2}{4c_0}\int_0^z e^{jk\psi}e^{-\alpha_2\psi}\left[\phi_1(r, z - \frac{\psi}{2})\right]^2 e^{-2\alpha_1(z-\psi)}d\psi \quad (14)$$

$\beta = 1 + \frac{1}{2}B/A$ et $\alpha_2$ représente l'atténuation du second harmonique.

3.2. Potentiel et pression acoustique moyenne du second harmonique

Le potentiel moyen s'exprime de la même façon que (3), et on obtient :

$$\left\langle \phi_2(r,z)\right\rangle \approx -\frac{\beta k^2}{4c_0}\int_0^z e^{jk\psi}e^{-\alpha_2\psi}\left\langle \phi_1\left(r, z-\frac{\psi}{2}\right)^2\right\rangle e^{-2\alpha_1(z-\psi)}d\psi \quad (15)$$

Nous la qualifierons de solution analytique de référence pour le potentiel moyen du second harmonique en milieu dissipatif, car les expressions que nous exploiterons par la suite seront des approximations de cette solution.
Ingenito et Williams [18] ont montré qu'une bonne approximation consistait à remplacer $<\phi_1^2>$ par $<\phi_1>^2$ dans l'expression de $<\phi_2>$. Ainsi, avec (5) et (12), on peut écrire :

$$\left\langle \phi_1(r,z)\right\rangle^2 = \left\langle \phi_{10}(r,z)\right\rangle^2\left[1-\left\{2g(z)-g(z)^2\right\}\right] = \left\langle \phi_{10}(r,z)\right\rangle^2\left[1-f(z)\right] \quad (16)$$

$$\text{où} \quad f(z) = 2g(z)-g(z)^2 = 1-D_1(z)^2 \quad (17)$$

Et comme $<p_2(r,z)> = -2j\rho_o\omega<\phi_2(r,z)>$, la solution (15) conduit à la pression moyenne :

$$\left\langle p_2(r,z)\right\rangle \approx \left\langle p_{20}(r,z)\right\rangle\left\{1-\frac{2\alpha_1-\alpha_2}{e^{(2\alpha_1-\alpha_2)z}-1}\int_0^z e^{-(\alpha_2-2\alpha_1)\psi}f\left(z-\frac{\psi}{2}\right)d\psi\right\} \quad (18)$$

$$\text{avec} \quad \left\langle p_{20}(r,z)\right\rangle = KP_0^2\left(\frac{e^{-\alpha_2 z}-e^{-2\alpha_1 z}}{2\alpha_1-\alpha_2}\right)e^{j2kz} \quad \text{et} \quad K=\frac{(2+B/A)\omega}{4\rho_0 c_0^3} \quad (19)$$

$<p_{20}(r,z)>$ représente la pression moyenne exercée par le second harmonique sur le détecteur dans le cas d'une onde plane, et l'expression (18) met ainsi en évidence la décroissance due à la diffraction par le second terme de { }. Avec (16) et (5) on obtient une forme équivalente mais plus compacte en fonction de $D_1(z)$ :

$$\left\langle p_2(r,z)\right\rangle \approx \left(KP_0^2 e^{-2\alpha_1 z}\int_0^z e^{(2\alpha_1-\alpha_2)\psi}D_1\left(z-\frac{\psi}{2}\right)^2 d\psi\right)e^{j2kz} \quad (20)$$

3.3. Fonction de correction de la diffraction pour le second harmonique

Comme pour le fondamental elle s'exprime sous la forme $D_2(z) = \frac{\left\langle p_2(r,z)\right\rangle}{\left\langle p_{2o}(r,z)\right\rangle}$, ainsi :

$$D_2(z) = 1-\frac{2\alpha_1-\alpha_2}{e^{(2\alpha_1-\alpha_2)z}-1}\int_0^z e^{(2\alpha_1-\alpha_2)\psi}f\left(z-\frac{\psi}{2}\right)d\psi \quad (21)$$

En fait, l'introduction d'une correction de la diffraction dans l'expression de la pression acoustique n'a d'intérêt par rapport à la solution (18) que si $D_2(z)$ est indépendante de l'atténuation. Cela revient à séparer les effets de l'atténuation et de la diffraction. En considérant $(\alpha_2 - 2\alpha_1)z$ faible, $D_2(z)$ se simplifie sous la forme:

$$D_2(z) = 1 - \frac{1}{z}\int_0^z f(z - \frac{\psi}{2})d\psi = \frac{1}{z}\int_0^z D_1\left(z - \frac{\psi}{2}\right)^2 d\psi \quad (22)$$

*Pour les liquides on a* $\alpha_2 = 4\alpha_1 = 4(\alpha_0 f^2)$*, et avec* $ka = 100$ *et* $z/a = 10$ *l'erreur sur* $|D_2(z)|$ *apportée par la simplification (22) est de 0.026% dans l'eau et 3.5% dans le glycérol, avec respectivement* $(\alpha_2 - 2\alpha_1)z = 0.028$ *et* $(\alpha_2 - 2\alpha_1)z = 4.8$*. Ces valeurs sont obtenues avec la solution exacte (7) pour* $D_1(z)$*.*
La définition d'une fonction de correction permet donc d'écrire la pression moyenne sous la forme : $\langle p_2(r,z)\rangle = \langle p_{20}(r,z)\rangle D_2(z)$ où les effets de l'atténuation sont intégrés à l'expression de $<p_{20}(r,z)>$.

## 4. Simplifications de $D_2(z)$ et $<p_2(r,z)>$

D'après (22) la correction $D_2(z)$ est fonction de $D_1(z)^2$ qui peut être simplifiée sous certaines conditions. Puisque $D_1(z)^2 = 1 - 2g(z) + g(z)^2$ et $\lim_{ka \to \infty}[g(z)] = 0$, nous pouvons négliger le terme $g(z)^2$ pour ka grand. Ainsi les solutions (11) et (9), valides pour $z/a < (ka)^{1/2}$ ou $s < 2\pi/(ka)^{1/2}$, avec $(ka)^{1/2} >> 1$, conduisent respectivement aux expressions simplifiées :

$$D_1(z)^2 \approx 1 - 2\left(\frac{2.z}{\pi ka^2}\right)^{1/2} e^{-j\pi/4} \quad (23)$$

$$D_1(z)^2 \approx 1 - 2\left(1 - \frac{\xi(z)^2}{2(ka)^2}\right)\left(\frac{2}{\pi\xi(z)}\right)^{\frac{1}{2}} e^{-j\frac{\pi}{4}} \quad (24)$$

Cette dernière expression a été utilisée par Coob [13] pour calculer la pression moyenne du second harmonique à l'aide d'une relation équivalente à notre formulation (20). Les mesures qu'il a réalisées à 3MHz dans l'eau (ka ≈78) et le glycérol (ka ≈60) pour $2 < z/a < 14$ [2] sont en très bonne concordance avec ses prédictions théoriques correspondant à la solution (20-24) qui ne peut être évaluée que par intégration numérique.

En conséquence, comme nous avons montré que les fonctions $D_1(z)$ (11) et (9) sont pratiquement équivalentes (figure 2), nous pouvons exploiter l'expression plus simple (23) avec une bonne précision sous la condition $z/a < (ka)^{1/2}$ avec $(ka)^{1/2} >> 1$.

Dans ce cas l'intégrale (22) peut être évaluée, et on obtient :

$$D_2(z) \cong 1 - C\sqrt{\frac{z}{ka^2}}.e^{-j\pi/4} \quad \text{avec} \quad C = \frac{4}{3\sqrt{\pi}}\left(2\sqrt{2} - 1\right) \approx 1.375 \quad (25)$$

Finalement, avec (19) nous pouvons établir une expression simple, mais suffisamment précise, permettant le calcul de la pression moyenne exercée par le second harmonique sur un récepteur de mêmes dimensions que la source :

$$|\langle p_2(r,z)\rangle| \approx KP_0^2\left(\frac{e^{-\alpha_2 z} - e^{-2\alpha_1 z}}{2\alpha_1 - \alpha_2}\right)|D_2(z)| \quad (26)$$

---

[2] *La borne inférieure est limitée expérimentalement à l'apparition d'ondes stationnaires.*

Cette relation est valable quelle que soit l'ampleur de l'atténuation.
En milieu moyennement dissipatif, et pour la plupart des milieux biologiques ($\alpha_2 \approx 2\alpha_1$), la quantité $(\alpha_2 - 2\alpha_1).z$ est faible et le terme d'atténuation entre ( ) dans l'expression (26) se simplifie pour donner :

$$\left|\langle p_2(r,z)\rangle\right| \approx K P_0{}^2 z e^{-\left(\alpha_1 + \frac{\alpha_2}{2}\right)z} \left|D_2(z)\right| \quad (27)$$

La simplification sur l'atténuation transformant la relation (26) en (27) introduit une erreur inférieure à 1% si $(\alpha_2 - 2\alpha_1)z < 0.5$.

D'autre part, nous remarquons que l'intégrale (22) peut être également évaluée en utilisant l'expression (11) au lieu de (23). Nous obtenons dans ce cas :

$$D_2(z) \approx 1 - C\sqrt{\frac{z}{ka^2}} e^{-j\pi/4} - j\frac{3}{2\pi}\frac{z}{ka^2} \quad (28)$$

Cependant, comme nous le verrons dans la partie suivante, cette expression est moins précise que la formulation (25). Notons également que l'intégrale (22) avec (9) au lieu de (24) ne peut être évaluée que par intégration numérique ou à l'aide de développements avec des fonctions de Bessel [13].

## 5. Comparaison des différentes solutions pour la pression moyenne $|\langle p_2(r,z)\rangle|$

Nous simulons les différentes expressions de la pression moyenne relative $|\langle p_2(r,z)\rangle|/Po$ dans deux milieux extrêmes sur le plan de l'atténuation et des effets non linéaires :
*Eau :* $c_0=1483$ *m/s,* $\rho_0=1000$ *kg/m*$^3$, $\alpha_0=0.25.10^{-13}$ *Npm*$^{-1}$*Hz*$^{-2}$, $\alpha_2=4.\alpha_1$, *B/A=5.2*
*Glycérol :* $c_0=1909$ *m/s,* $\rho_0=1260$ *kg/m*$^3$, $\alpha_0=26.10^{-13}$ *Npm*$^{-1}$*Hz*$^{-2}$, $\alpha_2=4.\alpha_1$, *B/A=9.4*

Les conditions, voisines des expérimentations de Coob, sont : f = 3 MHz, a = 1 cm, $I_0 = 0.5$ W/cm² pour l'eau et $I_0 = 10$ W/cm² pour le glycérol, avec $I_0 = P_0{}^2/(2\rho_0 c_0)$.

Les résultats sont présentés figure 3 et on constate que notre solution (26-25) est confondue avec celle de Coob (20-24), également confondue avec la simplification (27-25) dans un milieu faiblement dissipatif comme l'eau.
L'importance de la correction de la diffraction $D_2(z)$ est visualisée par la représentation de la solution (19) correspondant au cas simple d'une onde plane, c'est à dire pour $D_2(z) = 1$.
Nous avons également simulé la pression moyenne relative obtenue avec la solution de référence (15) et l'intégrale de King (2)[3] pour le fondamental $\phi_1$.

Les erreurs relatives entre la solution de référence et les solutions (26-25), (26-28) et (20-24), sont présentées figure 3.c-d pour l'eau et le glycérol, où les simulations sont effectuées avec une tolérance de $10^{-6}$ pour le calcul des intégrales avec le logiciel Mathcad. Elles montrent que la solution (26-25) est globalement plus précise que les solutions (20-24) et (26-28) dans les conditions adoptées pour la simulation. De plus, le temps de calcul nécessaire à la solution (26-25) est beaucoup plus faible que celui de la solution de référence qui comporte une intégrale triple.

---

[3] Nous avons exploité ici une forme plus adaptée que (2) établie par Hutchins et al [21]

## 6. Conclusion

Nous avons montré dans cet article que l'on pouvait obtenir une fonction de correction de la diffraction pour le second harmonique beaucoup plus simple que celles habituellement utilisées. Cette nouvelle formulation reposant sur une simplification de la correction à appliquer au fondamental de la pression acoustique.
L'expression obtenue conduit à des solutions simples pour le second harmonique de la pression détectée, mais suffisamment précises pour être exploitées dans les mesures du paramètre de non-linéarité B/A [22, 23]. Ces solutions, utilisables dans des cellules de mesure constituées de deux transducteurs plans de mêmes dimensions, sont valides pour $z/a < (ka)^{1/2}$ avec $(ka)^{1/2} >> 1$ et globalement plus précise que la solution de Coob vérifiée expérimentalement [14]. Un autre intérêt de ces solutions analytiques simples est l'importante réduction des temps de calcul lorsqu'elles sont intégrées dans des processus de simulation de systèmes travaillant dans le domaine de l'acoustique non linéaire [22].

## Bibliographie

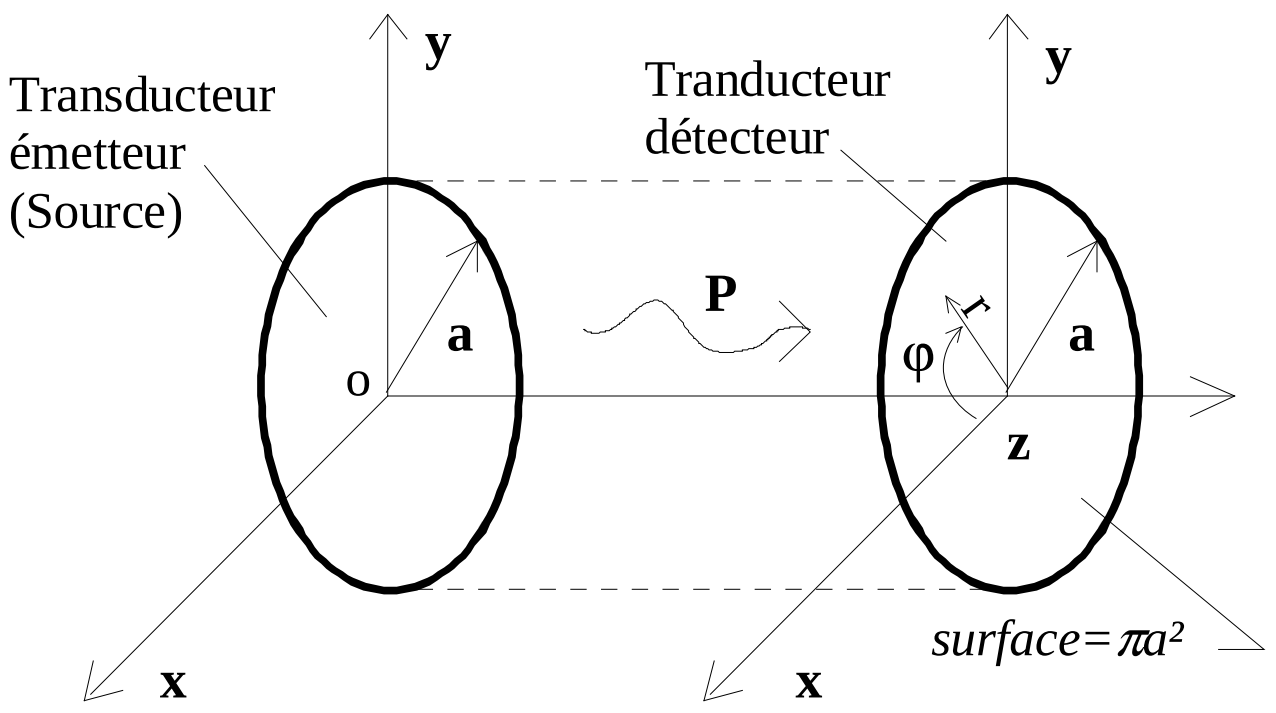

Figure 1. Configuration géométrique de l'ensemble émetteur-détecteur.

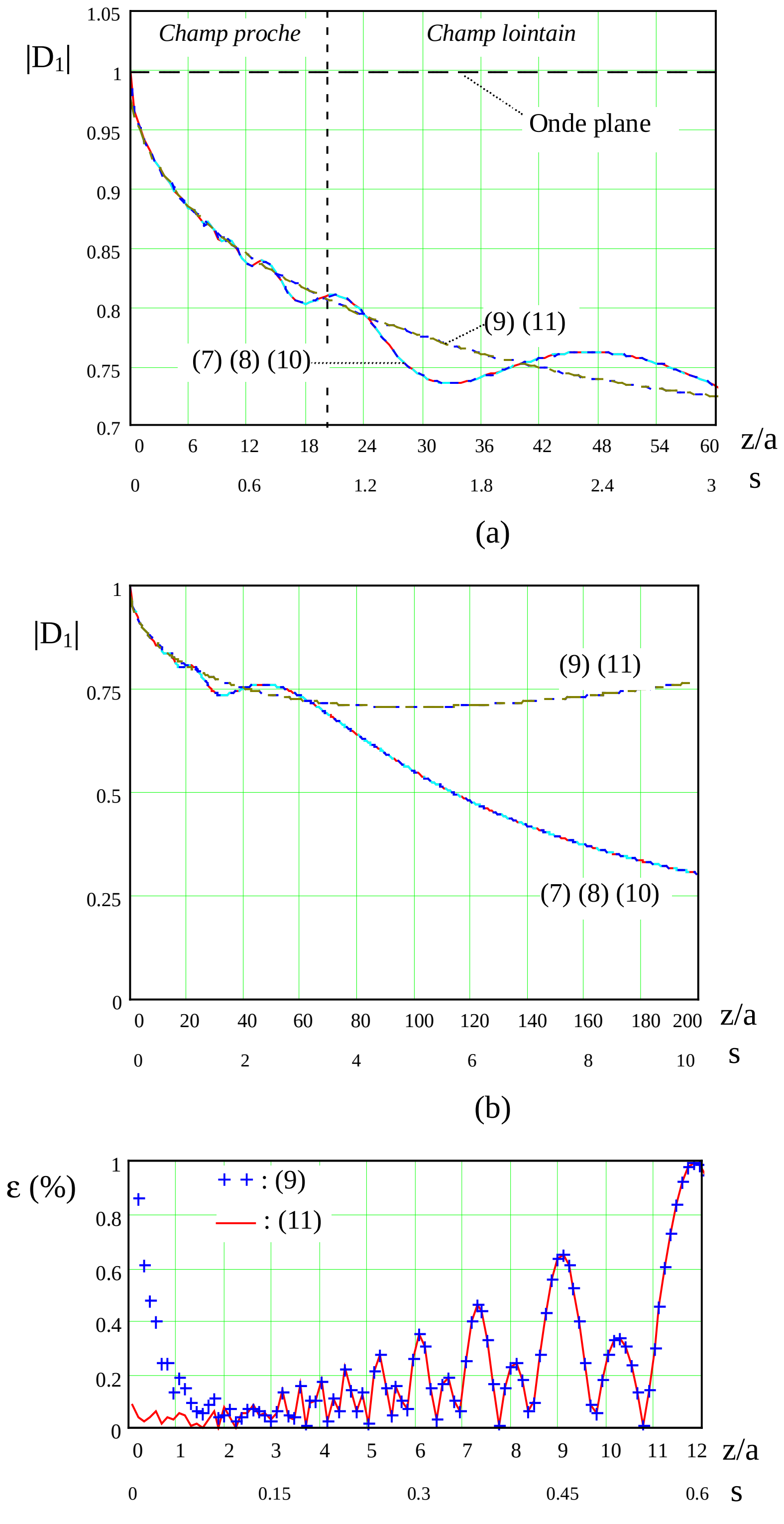

Figure 2. Fonctions de correction de la diffraction. Comparaison avec la solution exacte de Williams (7). La figure (c) présente l'erreur relative entre la solution exacte (7) et les solutions (9) et (11)

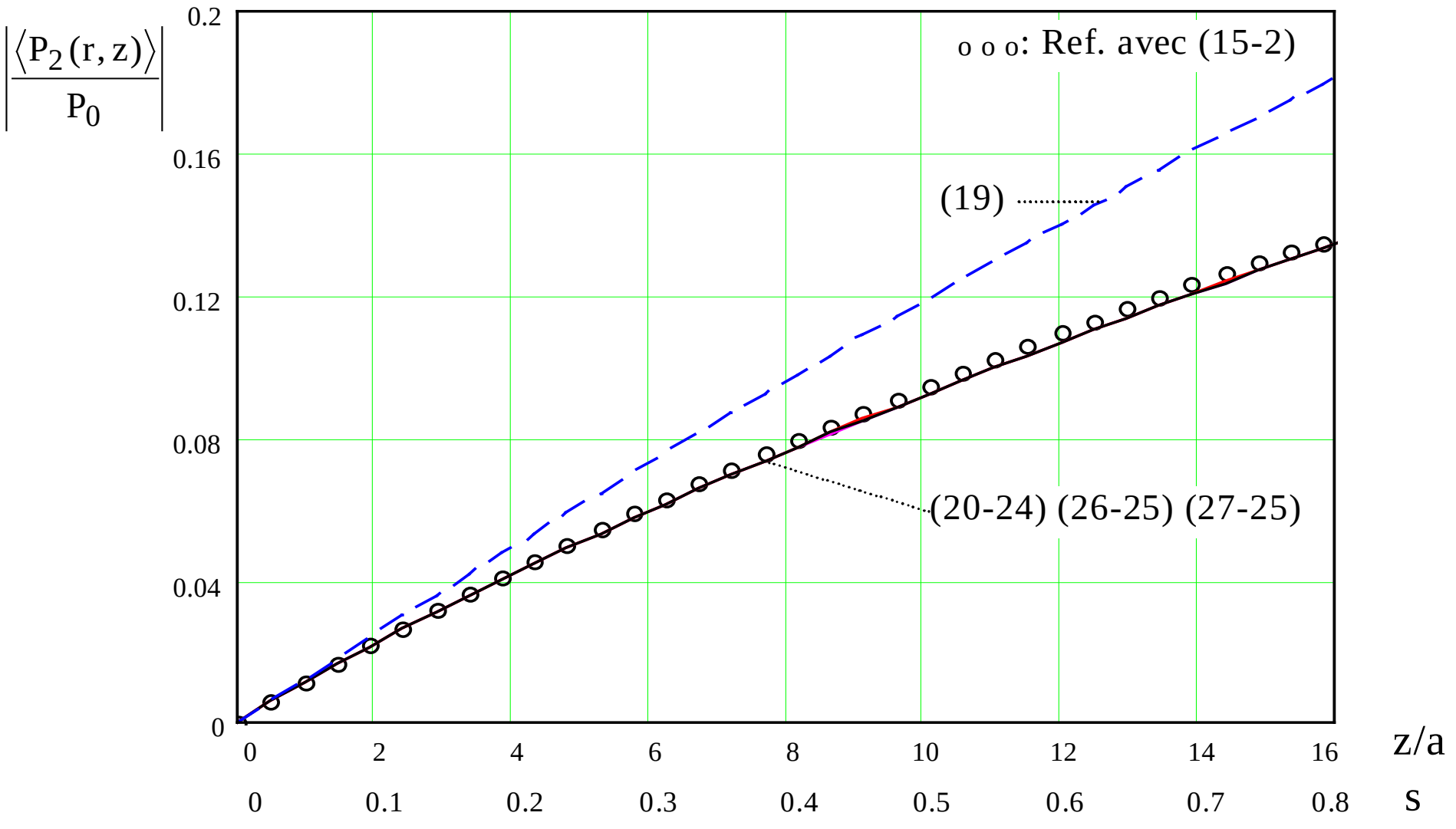

(a)

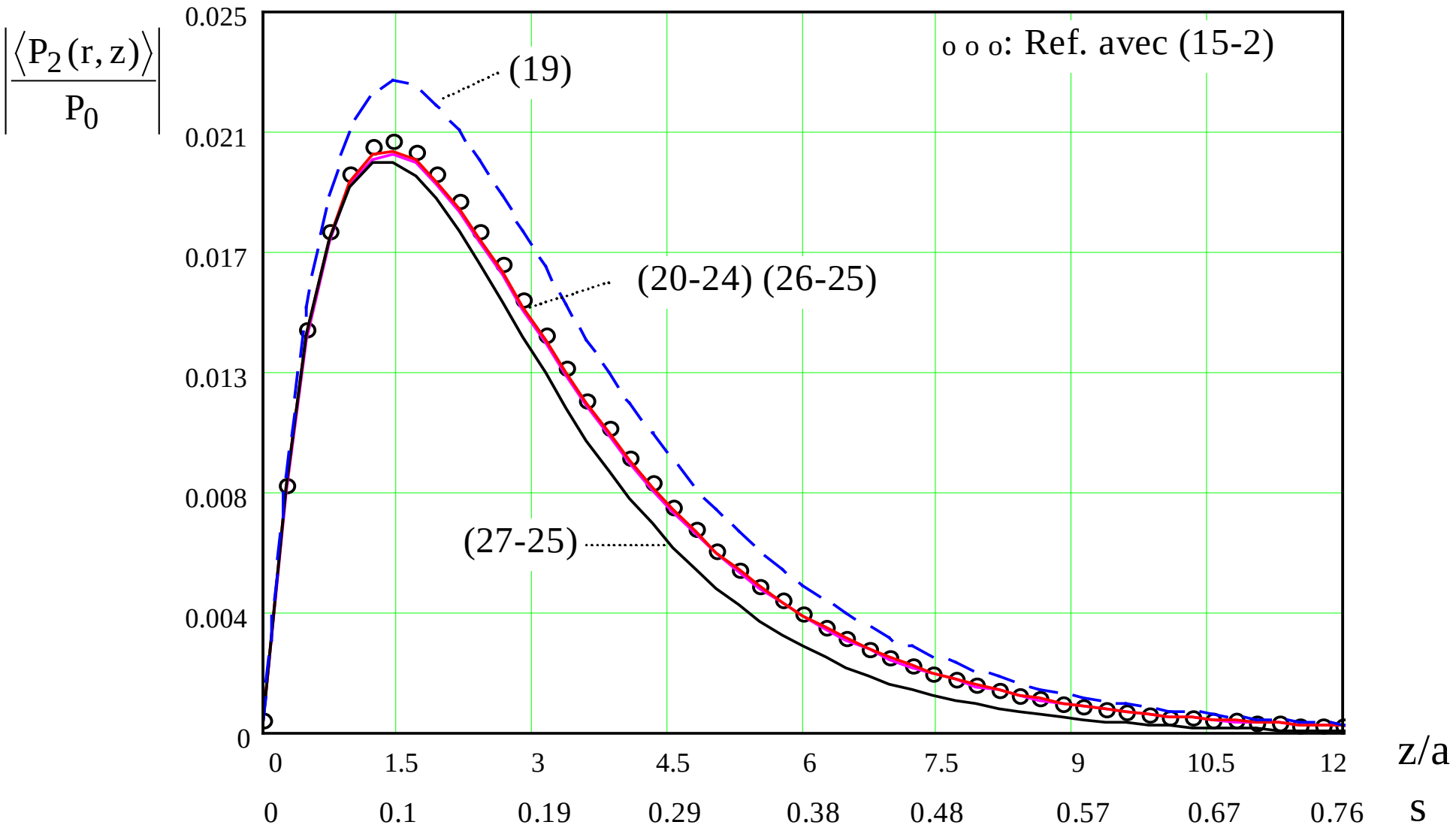

(b)

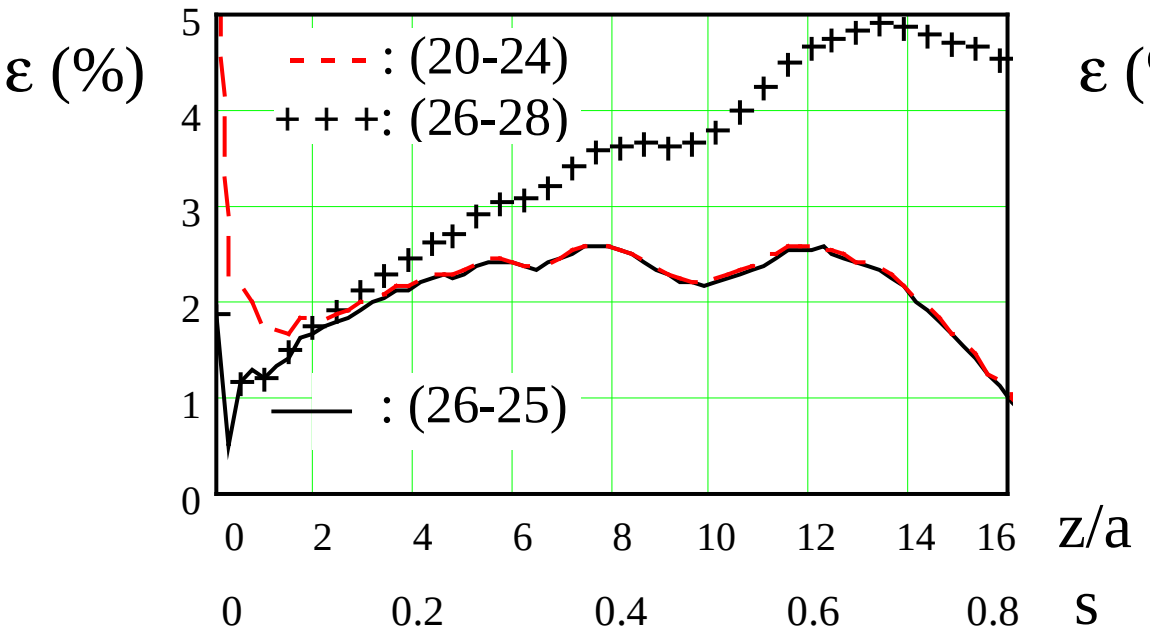

(c)

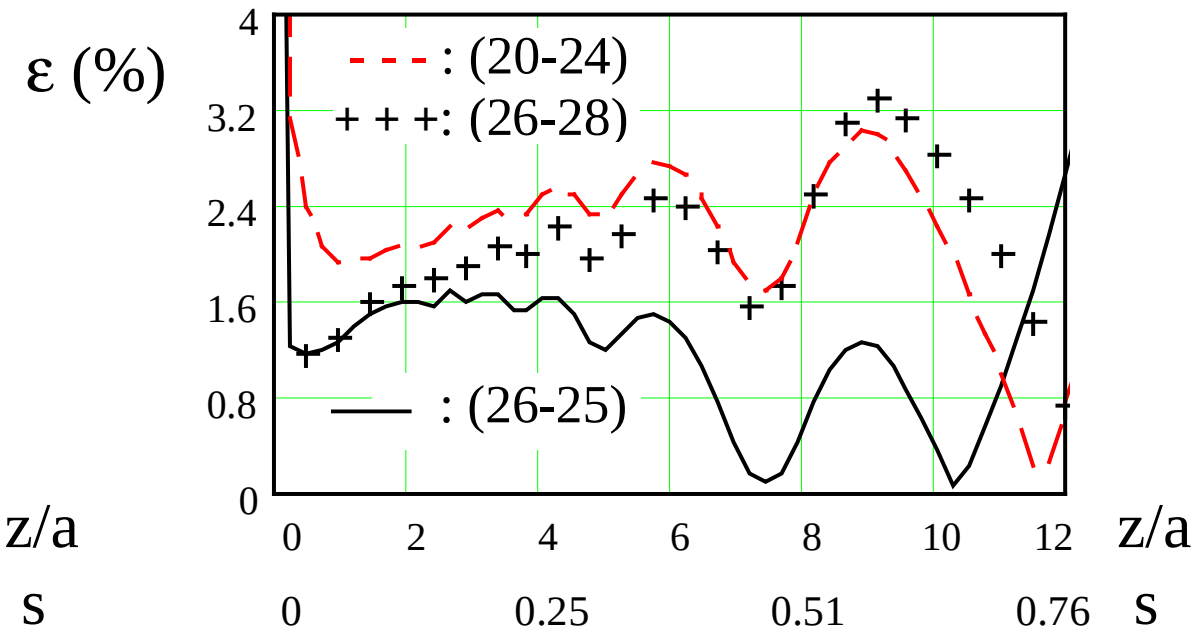

(d)

Figure 3. Simulations des différentes solutions analytiques de la pression moyenne du second harmonique pour l'eau (a) et le glycérol (b)
Ecart relatif entre la solution de référence (15-2) et les solutions (26-25), (20-24), et (26-28), pour l'eau (c) et le glycérol (d).